# Point contact Andreev reflection spectroscopy of superconducting energy gaps in 122-type family of iron pnictides

P. Samuely,[a] Z. Pribulová,[a] P. Szabó,[a] G. Pristáš,[a] S. L. Bud'ko,[b] P. C. Canfield[b]

[a] *Centre of Low Temperature Physics, Institute of Experimental Physics, Slovak Academy of Sciences, Watsonova 47, Košice, Slovakia*
[b] *Ames Laboratory, Iowa State University, Ames, Iowa 50011, USA*

**Abstract**

A brief overview of the superconducting energy gap studies on 122-type family of iron pnictides is given. It seems that the situation in the hole-doped $Ba_{1-x}K_xFe_2As_2$ is well resolved. Most of the measurements including the presented here point-contact Andreev reflection spectra agree on existence of multiple nodeless gaps in the excitation spectrum of this multiband system. The gaps have basically two sizes – the small one with a strength up to the BCS weak coupling limit and the large one with a very strong coupling with $2\Delta_L/kT_c > 6 - 8$. In the electron doped $Ba(Fe_{1-x}Co_x)_2As_2$ the most of the experiments including our point contact measurements reveal in quite broadened spectra only a single gap with a strong coupling strength. The high precision ARPES measurements on this system identified two gaps but very close to each other, both showing a strong coupling with $2\Delta/kT_c \sim 5$ and 6, respectively.

## 1. Introduction

Iron pnicitides [1] – finally a new family of high-$T_c$ superconductors - represent a real challenge in the recent condensed matter physics. Despite an enormous effort during less than one year after their appearance many puzzles remain unsolved. One of the important questions concerns the superconducting order parameter in these systems with a strongly multiband character. In this paper a brief overview of the experimental studies of the order parameter in the 122-type family of iron pnictides is given.

Similarly to the high-$T_c$ cuprates the superconductivity in iron pnictides is enabled by chemical doping of the antiferromagnetic parental compounds which in contrast to the cuprates are metallic. The highest transition temperature (up to 56 K) among different iron pnicitides has been achieved in the optimally doped REFeAsO(F), or the 1111 group with Gd, Nd, or Sm [2] standing for a rare element RE. Considerable interest has also been attracted by another class of iron pnictide superconductors based on $AFe_2As_2$ with $A$ = Ba, Sr and Ca, referred to as the 122-type group. The 122-type compounds are chemically and structurally simpler and less anisotropic than the 1111 ones. The maximum $T_c$ of 38 K is obtained in the optimally hole doped $Ba_{0.6}K_{0.4}Fe_2As_2$ system [3] but also the electron doped $Ba(Fe_{1-x}Co_x)_2As_2$ crystals with $T_c$ about 25 K are available [4]. In contrast to the 1111 systems the 122-type parent compounds show magnetic (from paramagnetic to SDW antiferromagnetic phase) and structural transition (from tetragonal to orthorhombic phase) at the same temperature of about 140 K. This transition is gradually suppressed by chemical doping but the phase diagram *Temperature versus doping* shows an overlap between the SDW/orthorhombic and superconducting phases for x=0.2 to 0.4 in Ba$_{1-x}$K$_x$Fe$_2$As$_2$ [5]. Band structure calculations [6] have shown that the Fe *3d* bands located near the Fermi energy are responsible for the appearance of multiple Fermi surface (FS) sheets. The multiband/multigap superconductivity with interband interactions leading to an exotic *s*-wave pairing with a sign reversal of the order parameter between different FS sheets [7] stands among the hot candidates to explain the high-$T_c$ superconductivity in iron pnictides.

The following brief overview of some of the available experimental data brings pieces of evidence for multiple superconducting energy gaps revealing basically two sizes of the coupling strength $2\Delta/kT_c$. The most of the data point to an *s*-wave pairing symmetry.

## 2. Overview of the superconducting gap studies

High resolution angle resolved photoemission spectroscopy (ARPES) has become extremely effective tool for studies of the FS sheets and superconducting energy gaps in a momentum space. Ding *et al.* [8] observed three FS sheets in Ba$_{0.6}$K$_{0.4}$Fe$_2$As$_2$ crystals with $T_c$ of 37 K: an inner hole-like FS pocket, an outer hole-like Fermi surface sheet, both centered at the Brillouin zone center Γ and a small electron-like FS at the M point. A large superconducting energy gap ($\Delta_L$ = 12 meV) was detected on the two small hole-like and electron-like FS sheets while a small gap ($\Delta_S$ = 6 meV) was found on the large hole-like FS. The gaps closing at the same $T_c$ are isotropic. Two small FS sheets with a very strong coupling strength $2\Delta_L/kT_c\sim 8$ are connected by the (π,0) SDW vector in the parent compound indicating an importance of the interband interaction between these two nested FS sheets also for superconductivity. Similar results were obtained also in Ref. 9, 10 and 11. Wray *et al.* [11] proposed that the observed gap structure oscillating among the FS sheets is consistent with an order parameter that takes the in-plane form of $\Delta_0 cos(k_x)cos(k_y)$. Nakayama *et al.* [12] reported on the observation of the fourth FS sheet in Ba$_{0.6}$K$_{0.4}$Fe$_2$As$_2$. It is another, outer electron pocket centered around the M point with the superconducting energy gap comparable (~11 meV) to the gap on the inner electron and hole pockets.

Important finding has been reported by Terashima *et al.* [13] on ARPES measurements on Ba(Fe$_{1-x}$Co$_x$)$_2$As$_2$ crystals. In the sample with x=0.15 and $T_c$ of 25 K due to the electron doping the inner hole-like FS sheets is absent and the nesting conditions are switched from the inner hole FS sheet to the outer one which is connected to the electron FS sheets by the (π,0) SDW vector. Strong coupling strengths $2\Delta/kT_c\sim 6$ and $2\Delta/kT_c\sim 4.5$ are found on the hole and electron FS's, respectively. In heavy Co doped samples only the electron-like FS sheets remain and no superconductivity is present. All this is supporting the inter-FS superconductivity in iron pnictides.

Two-gap nature in Ba$_{1-x}$K$_x$Fe$_2$As$_2$ is supported also by infrared spectroscopy experiments [14] and specific heat measurements [15]. The penetration depth, or the lower critical field measurements by various techniques suggest a multigap picture in 122 as well. In Ba$_{1-x}$K$_x$Fe$_2$As$_2$ Hashimoto *et al.* [16] found the exponential temperature dependence of the superfluid density compatible with the fully opened two gaps. Particularly interesting was that only in the cleanest crystals this effect was detectable. Similar results are presented also in Ref. 17. Vorontsov *et al.* [18] argue that also the penetration depth measurements in another isotypic structure of Ba(Fe$_{1-x}$Co$_x$)$_2$As$_2$ can be explained by a two gap scenario with extended *s*-wave pairing.

Only few tunneling spectroscopy measurements are available so far generally in iron pnictides. Scanning tunneling microscope (STM) measurements were performed on the 32 K Sr$_{1-x}$K$_x$Fe$_2$As$_2$ [19]. On the surfaces with a stripe-like modulation on a square atomic lattice consisting of either Sr/K or As sometimes a gapped spectrum is observed with coherent peaks at 10 mV. This would correspond to the coupling strength $2\Delta/kT_c\sim 7$. Yin *et al.* [20] in their STM measurements on the Ba(Fe$_{0.9}$Co$_{0.1}$)$_2$As$_2$ crystals found a relatively small variation of the well pronounced gap with the averaged $\Delta\sim 6$ meV, corresponding to $2\Delta/kT_c\sim 6$ (but the local $T_c$ was not established). Also disordered vortices were detected showing neither localized states typical for an *s*-wave superconductors in the clean limit nor any internal structure predicted for *d*-wave vortices. Much bigger spatial variations from the averaged gap value were reported by Massee *et al.* [21] in their STM measurements on Ba(Fe$_{0.93}$Co$_{0.07}$)$_2$As$_2$ with $2\Delta/kT_c$ spanning between 5 and 10.

## 3. Point contact Andreev reflection spectroscopy on 122-type iron pnictides

Point contact Andreev reflection (PCAR) spectra measured on the ballistic microconstriction between a normal metal and a superconductor consists of pure Andreev reflection and tunneling contributions, respectively [22]. First contribution makes the conductance inside the voltage region /V/

< $\Delta/e$ twice as large as in the normal state or as what is at large bias where the coupling via the gap is inefficient. Tunneling contribution reduces the conductance at the zero bias and two symmetrically located peaks rise at the gap energy. PCAR conductance can be compared with the Blonder-Tinkham-Klapwijk (BTK) model using as input parameters the energy gap $\Delta$, the parameter $z$ (measure for the strength of the interface barrier) and a parameter $\Gamma$ for the spectral broadening. For a multiband/multigap superconductor the point contact conductance $dI/dV$ can be expressed as a weighted sum of partial BTK conductances. PCAR spectroscopy has proved to be very powerful in the investigation of the two gap superconductor $MgB_2$ [23]. In $MgB_2$ the total PCAR conductance consists of two parallel contributions, the first originated from a three dimensional $\pi$ band with a small gap $\Delta_S$ and a weight $\alpha$ and the second from a quasi two dimensional $\sigma$ band with a large gap $\Delta_L$ and a weight $(1-\alpha)$, respectively. In that case, the parameter $\alpha$ varied between 0.7 and 0.99, depending on angle between the point contact current and orientation of $MgB_2$ crystal. Also different values of $z$ for each band were necessary in the fitting procedure.

Spectral characteristics as the superconducting energy gaps or electron-phonon interaction features can be read from the point contact data only if the junction is in a ballistic or diffusive regime, where heating effects are avoided. There, both the elastic $l_e$ as well as inelastic $l_i$ mean free paths should be larger than the diameter $d$ of the junction or a diffusion length $(l_e.l_i)^{1/2} > d$. These requirements should be satisfied in the normal as well as in the superconducting part of the junction.

$MgB_2$ was quite an exceptional example where the PCAR spectroscopy worked extremely well but the situation in iron pnictides is far more complicated. The latter systems are highly resistive materials where reaching of ballistic regime is quite challenging. Point contacts (PC) made on a highly resistive material as $Ba_{1-x}K_xFe_2As_2$ have resistances between tens and hundreds of ohms which corresponds to the contact diameter of tens of nanometers. Indeed, the electron mean free path can also be on the same order, here. Then, precautions should be made to avoid the junctions with heating effects. In our work on iron pnictides only the junctions without the conductance dips and irreversibilities in voltage dependences are presented. Moreover to preserve the spectroscopic conditions we focus on PC junctions having a finite barrier strength parameter $z$ with a tunneling component in the spectrum.

A co-existence of phase separated regions with a static magnetic order and superconducting islands was demonstrated in single crystalline $Ba_{1-x}K_xFe_2As_2$ [24,25]. The size of the phase separated regions of about 50 nm is again comparable with the PC diameter, a region mostly contributing to the PCAR spectrum. This must be taken into considerations to avoid a misinterpretation of the observed data.

In the following we review our point contact spectroscopy study on the $Ba_{0.55}K_{0.45}Fe_2As_2$ single crystals recently published in Ref. 26. The crystals were grown out of a Sn flux. The resistive measurements showed the onset of the superconducting transitions below 30 K and the zero resistance at 27 K. Rather broad transitions in some of the crystals with multiple steps are attributed to a possible different amount of potassium in different layers or different crystals. A local transition temperature measured by the point-contact technique showed superconducting $T_c$'s between 23 and 27 K.

The specific heat as well as the resistivity measurements on these crystals showed features at about 85 K. Although reduced they are found at the same temperature as on the undoped $BaFe_2As_2$ samples where the tetragonal-to-orthorhombic structural phase transition takes place. This transition is revealed here at a lower temperature as compared with 140 K found in Ref. 3. The decreased structural transition temperature is due to the amounts of Sn up to 1% incorporated into the bulk. A presence of tin does not significantly effect the high-temperature superconducting phase transition in $Ba_{0.55}K_{0.45}Fe_2As_2$.

Before point contact measurements the crystals were cleaved to reveal fresh surface. For the measurements in the $c$ direction the fresh shiny surface was obtained by detaching the degraded surface layers by a scotch tape. The microconstrictions were prepared in situ by pressing a metallic tip (platinum wire formed either mechanically or by electrochemical etching) on a fresh surface of the superconductor. For the measurements with the point contact current in the *ab* plane a reversed tip-sample configuration was used. The freshly cleaved edge of the single crystal jetting out in *ab* direction was pressed on a piece of chemically etched copper. A special PC approaching system allowed for lateral as well as vertical movements of the PC tip by a differential screw mechanism. Details of the technique can be found elsewhere [23].

Figure 1 shows typical PCAR spectra obtained on $Ba_{0.55}K_{0.45}Fe_2As_2$ with the PC current preferably within the *ab* plane. The spectra present double enhanced conductances, the typical features of the

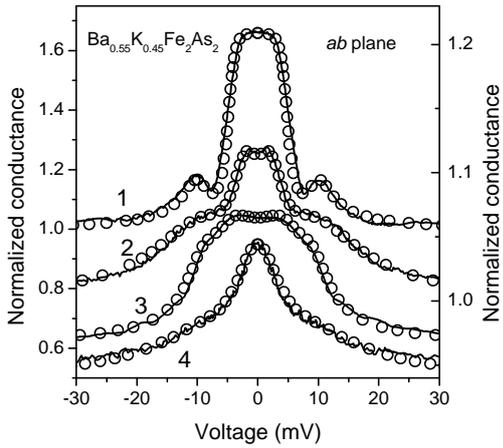

Fig.1 Typical PCAR spectra on $Ba_{0.55}K_{0.45}Fe_2$ (lines) measured at 4.4 K with the PC current mostly within the *ab* plane of the sample. Spectra are normalized to their respective normal state and fitted to the two gap BTK model (symbols). For curves 2 and 4 right *y* axis applies. Spectra 2, 3 and 4 are vertically shifted for clarity.

Andreev reflection of quasiparticles coupled via two superconducting energy gaps. The first enhancement starts below 20 mV with the gap-like humps at about 10 mV while the second one is located below ~ 5-7 mV. On the spectra 2 and 3 also two symmetrical maxima at 2-3 mV are displayed. Majority of the spectra measured in the *ab* direction revealed a heavily broadened enhanced conductance near the zero bias as indicated by the spectrum 4. This is most probably caused by the sample inhomogeneities on the nanoscale.

The presented spectra are normalized to their respective normal state and fitted to the two gap BTK model (symbols). The resulting values of energy gaps are spread in the range of 2-5 meV and 9-10meV for the small and large gap, respectively. The values of smearing parameters were 10, 60, 30, and 100% of each energy gap value for curves 1, 2, 3, and 4, respectively. For each presented fit different values of $z$ for the two bands were also necessary. Typically, $z_L$ for the band with a large gap was about 0.4 – 0.8, while $z_S$ was twice smaller. Parameter α varied between 0.4 and 0.8. Although the s-wave two gap BTK formula has been successfully used to fit our PCAR data a possibility of unconventional pairing symmetry cannot be completely ruled out. Obviously, rather strongly broadened spectra as presented here could be in principle fitted also by model taking into account anisotropic or nodal gaps, if an appropriate current injecting angle was selected [27].

In Fig. 2a temperature evolution of the second spectrum from previous figure is presented. All the spectra (lines) were normalized to the conductance measured at 27 K and fitted to the BTK model with a proper temperature smearing involved. Obviously, the spectrum at the lowest temperature reminds the two gap spectrum of $MgB_2$ for a highly transparent junction with conductance enhancements due to Andreev reflection of quasiparticles. As the temperature is increased the double enhanced point contact conductance corresponding to two energy gaps is gradually smeared out and spectrum intensity decreases. Indeed, the spectra could be well fitted to the two gap BTK formula. The best fit for each temperature is shown by open circles. The extracted values of the gaps at different temperatures are shown in Fig. 2b (symbols) following nicely a BCS prediction (lines) rescaled to the size of the respective gap. The values of the energy gaps at lowest temperature are for the small $\Delta_S \sim 2.7$ meV and the large one $\Delta_L \sim 9.2$ meV, which corresponds to the coupling strengths $2\Delta_S/kT_c \sim 2.7$ and $2\Delta_L/kT_c \sim 9$ for $T_c$=23 K. The smearing parameters (about 60% of the respective gap values), the barrier strengths $z \sim 0.3$ and 0.6 as well as the weight factor $\alpha \sim 0.5$ obtained at 4.4 K were kept constant at higher temperatures. From the data obtained on more junctions we observe that the gaps are scattered as $2\Delta_S/kT_c \sim$ 2.5-4 and $2\Delta_L/kT_c \sim$ 9-10.

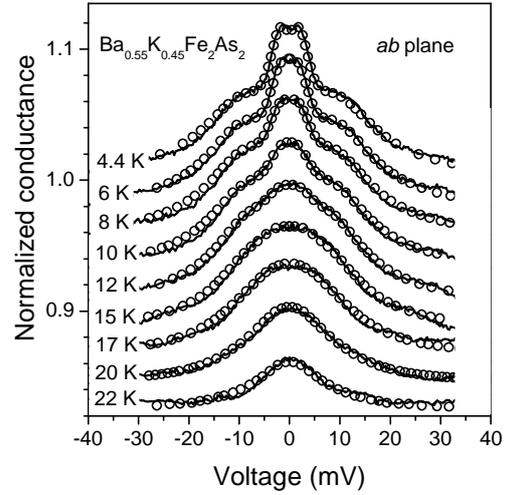

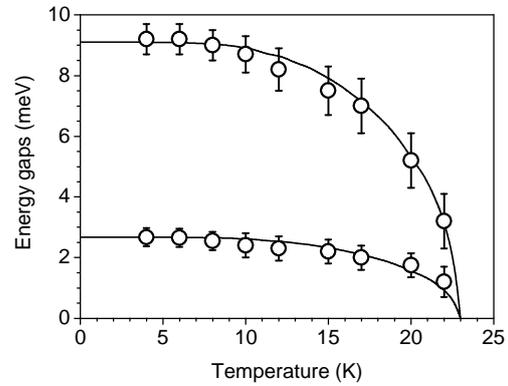

Fig.2. (a) Temperature evolution of spectrum 2 from Fig.1 (lines). Fits to the BTK model are indicated by symbols (b) Values of the energy gaps obtained from fitting procedure for distinct temperatures. Lines represent the BCS type temperature dependences of energy gaps.

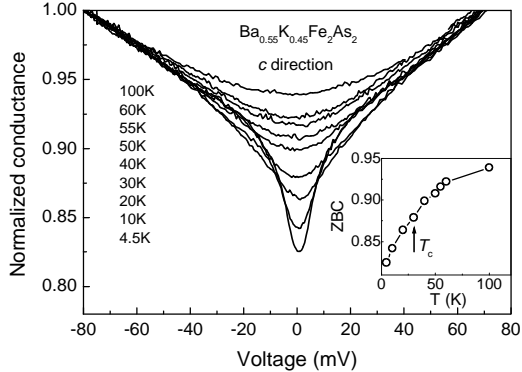

Fig.3. Spectrum of Pt-Ba$_{0.55}$K$_{0.45}$Fe$_2$ junction with PC current in the c direction showing a reduced conductance even above $T_c$. Inset – evolution of zero bias conductance with the temperature, arrow depicts a position of the critical temperature.

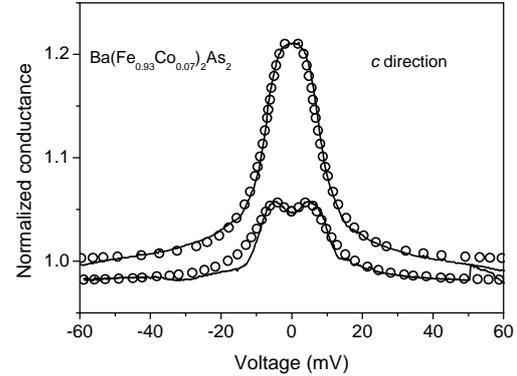

Fig.4. Spectra of Pt- Ba(Fe$_{0.93}$Co$_{0.07}$)$_2$As$_2$ junctions measured at 4.5 K (lines). Fits to the single gap BTK model are shown by circles.

The measurements on the Ba$_{0.55}$K$_{0.45}$Fe$_2$As$_2$ single crystals with the PC current in the *c* direction yield completely different picture showing just a reduced conductance around zero bias. In Fig. 3 the temperature dependence of such a spectrum is displayed. The zero-bias conductance minimum is step by step smeared and filled up with increasing temperature. However, the filling effect, which cannot be explained just by the spectral broadening by temperature, is not finished at $T_c$ but continues up to about 70 - 80 K, the temperature at which the magnetic transition in the system takes place [28]. Thus, this feature is not related to superconductivity.

Absence of apparent superconducting gap features in the *c* axis junctions characteristics is probably caused by the surface contamination and reconstruction.

Remarkably a similar reduced conductance background around the zero bias is displayed sometimes also on the *ab* plane junctions revealing the two gaps at low temperatures [26]. The reduced conductance is then detectable near $T_c$ and persists again well above $T_c$ similarly to the case of the *c* axis spectra. This indicates that the reduced conductance is a spectral feature related to a reduced DOS in the normal state. A pseudogap in the quasiparticle excitation spectrum responsible for superconductivity is one of possible interpretations. But since there is an evidence for a mesoscopic phase separation of antiferromagnetically ordered and non-magnetic/superconducting regions in the similar single crystals [25] another possibility would be that there are parallel PC currents to these two regions. A rather large broadening of the PCAR spectral features revealed by $\Gamma$ parameters could also be related to the pair breaking effect originaning from static magnetic moments.

Recently we extended our PCAR measurements to the hole doped system of the 122-type family, namely to the optimally doped Ba(Fe$_{0.93}$Co$_{0.07}$)$_2$As$_2$ samples with $T_c$ of approximately 23 K determined from transport and susceptibility measurements. The single crystals were grown from FeAs/CoAs flux from a starting load of Ba, FeAs and CoAs precursors. Details of the preparation resulting to large crystals of a few mm size can be found elsewhere [29].

In Fig. 4 the typical spectra of the Pt-Ba(Fe$_{0.93}$Co$_{0.07}$)$_2$As$_2$ junctions measured at 4.5 K are presented. The spectra show enhanced differential conductances and the bottom curve reveal also a single pair of peaks at about 5 mV. In no case a double enhanced conductance spectrum indicative for two gap superconductivity was found. Indeed, the spectra on Ba(Fe$_{0.93}$Co$_{0.07}$)$_2$As$_2$ can be fitted just by a single *s*-wave gap BTK formula. This is in strong contrast to the PCAR spectra of the hole doped Ba$_{0.55}$K$_{0.45}$Fe$_2$As$_2$. The fits on the electron doped system have yielded the superconducting energy gap of around 5-6 meV. In all spectra a significant broadening was observed which is witnessed by a large value of the $\Gamma$ parameter spreading between 50% and 100% of the respective gap value. The superconducting transition temperature at the junctions was 22 K very close to the bulk $T_c$ determined from transport measurements. With $T_c$ the coupling strength $2\Delta/kT_c$ between 5.3 and 6.3 can be determined. This is remarkably consistent with the above mentioned results obtained by the ARPES measurements. Although our results are only preliminary and the statistics is not yet sufficient, the data seem to prove that if there are multiple gaps in the electron doped $Ba(Fe_{0.93}Co_{0.07})_2As_2$ superconductors they are quite close to each other.

*Note:* We are not aware of any other PCAR measurements done on 122-type iron pnictides so far but it is worth to note that on the 1111 superconductors similar results to ours were obtained by several groups. For example on LaFeAs(O,F) Gonnelli *et al.* [30] observed very pronounced two gap PCAR spectra. Their size of the large and small gaps on the samples with a similar $T_c$ are remarkably alike to those presented here.

**Conclusion**

In contrast to cuprates which are single band and *d* wave superconductors, 122-type iron pnictides seem to be proven as the multiband systems with multiple nodeless gaps. In the hole doped $Ba_{1-x}K_xFe_2As_2$ the available data point to an existence of two distinct superconducting energy gaps with the strength below and much above the single band BCS weak coupling limit, respectively. In the electron doped $Ba(Fe_{1-x}Co_x)_2As_2$ if there are two gaps present they are very close to each other having a strong coupling $2\Delta/kT_c$ between 5 and 6. Further experimental and theoretical effort is certainly needed to elucidate physics in these extremely interesting systems.

**Acknowledgement**

This work was supported by the Slovak R&D Agency under Contracts Nos. VVCE-0058-07, APVV-0346-07, and LPP-0101-06, and by the U.S. Steel Košice. Centre of Low Temperature Physics is operated as the Centre of Excellence of the Slovak Academy of Sciences. Work at the Ames Laboratory was supported by the U.S. Department of Energy, Basic Energy Sciences, under Contract No. DE-AC02-07CH11358. Valuable discussions with I.I. Mazin, A.A. Golubov and N.L. Wang are appreciated.